\documentclass[acmtog]{acmart}

\AtBeginDocument{%
  }

\setcopyright{rightsretained}
\copyrightyear{2023}
\acmDOI{2310.04080}

\usepackage{import}

\DeclareMathOperator*{\argmin}{arg\,min}

\definecolor{jkcol}{rgb}{0.3,0.7,0.5}

\definecolor{ktcol}{rgb}{0.7,0.3,0.2}

\begin{document}

\title{Robust Average Networks for Monte Carlo Denoising}

\author{Javor Kalojanov}
\email{jkalojanov@wetafx.co.nz}
\affiliation{%
  \institution{W\=et\=aFX}
  \country{New Zealand}
  \city{Wellington}
}
\author{Kimball Thurston}
\email{kthurston@wetafx.co.nz}
\affiliation{%
  \institution{W\=et\=aFX}
  \country{New Zealand}
  \city{Wellington}
}

\renewcommand{\shortauthors}{Kalojanov and Thurston}
\begin{abstract}
  We present a method for converting denoising neural networks from spatial into spatio-temporal ones by modifying the network architecture and loss function.
  We insert Robust Average blocks at arbitrary depths in the network graph.
  Each block performs latent space interpolation with trainable weights and works on the sequence of image representations from the preceding spatial components of the network.
  % These connections are kept live during training by pre-warping and modifying the input frames to improve feature coherence in space and time.
  The temporal connections are kept live during training by forcing the network to predict a denoised frame from subsets of the input sequence.
  Using temporal coherence for denoising improves image quality and reduces temporal flickering independent of scene or image complexity.
  % By increasing the amount of temporal information used for denoising, we reduce noise and flickering without sacrificing image quality by blurring in space.
  % We evaluate the resulting networks on complex renders of photorealistic scenes.
\end{abstract}

\begin{CCSXML}
  <ccs2012>
     <concept>
         <concept_id>10010147.10010371.10010372.10010374</concept_id>
         <concept_desc>Computing methodologies~Ray tracing</concept_desc>
         <concept_significance>500</concept_significance>
         </concept>
     <concept>
         <concept_id>10010147.10010371.10010382.10010383</concept_id>
         <concept_desc>Computing methodologies~Image processing</concept_desc>
         <concept_significance>500</concept_significance>
         </concept>
   </ccs2012>
\end{CCSXML}

\ccsdesc[500]{Computing methodologies~Ray tracing}
\ccsdesc[500]{Computing methodologies~Image processing}

\keywords{denoising, neural networks, ray tracing, monte carlo}
%   \def\svgwidth{\textwidth} 
%   
%   \import{}{teaser.pdf_tex}
%   % \includegraphics[width=\textwidth]{resnet.pdf}
%   %\caption{Noisy and denoised frame, and the magnitude of the temporal denoising weights using our method.}
%   \Description{teaser}

%\received{23 May 2023}
%\received[revised]{12 March 2023}
%\received[accepted]{5 June 2023}

\maketitle

\section{Introduction}

Deep Convolutional Neural Networks (CNNs) have been recently established as state-of-the-art techniques for image filtering in various areas.
In this paper, we discuss denoising images generated via Monte Carlo path tracing in Visual Effects (VFX) production.
Informally, denoising techniques reduce error by averaging pixel values in space and time, and the challenging aspects are balancing noise reduction with loss of detail due to blurring across image features.
The main benefit of using CNNs for denoising is the large amount of filters with trainable weights that can provide very good feature preservation.
This is especially important for VFX, where visual realism and image fidelity is often driven by the amount of detail in the rendered images.
In this paper, we propose a method that improves temporal averaging while preserving image detail.
We show that the technique applies to denoising complex imagery that contains interactions between rendered elements such as skin, water and hair.

Our method is targeted for production, where flickering, either introduced or left untreated by the denoising method, is not acceptable.
We incorporate temporal information into our neural networks via recurrent components that interpolate between the latent representations of consecutive frames.
These components consider a sequence of preceding and succeeding frames and convert spatial kernel-predictive neural networks into spatio-temporal ones.
The main benefits of kernel-predictive networks for our target application are the robustness against color shifts, and the ability to reuse denoising weights across multiple image buffers (AOVs).
We furthermore train without ground truth reference images, which can be prohibitively expensive to generate in large enough quantities in a production setting.

The contributions of this paper revolve around a set of convolutional network layers that we call Robust Average (RA) blocks.
These blocks perform interpolation and outlier removal in the time dimension.
They operate on the network's internal state and can be inserted at arbitrary depth.
The resulting networks have increased (but fixed) receptive field in time and better incorporate information from the entire frame sequence.
The balance between spatial and temporal information used for denoising is learned from the data.
This can lead to the networks degenerating to spatial ones during training.
To prevent this, we warp each frame inside the temporal window of the network to convert it, as much as possible, to an estimate of the frame being denoised (the central frame).
In order to further increase temporal contributions in the absence of temporal references during training, we propose a loss reformulation which forces the network to estimate the output frame from subsets of the input sequence that exclude the central frame.
We thereby convert spatial loss functions into temporal ones without losing the ability to penalize warping artifacts.

We also propose to use thresholds on the activations of the kernel predictive layers.
On one hand, this allows for user control of the influence of the denoising network on the input images.
On the other, it prevents artifacts created by overestimated kernel weights in high-intensity regions of the image.

\begin{figure*}
  \centering
  \includegraphics[width=\textwidth]{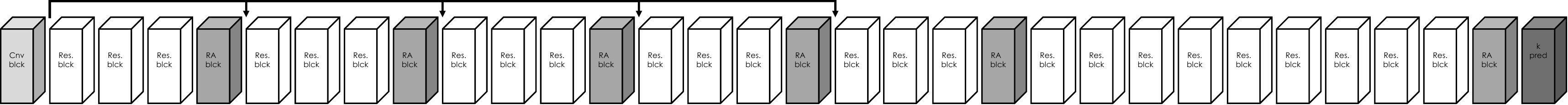}
  \caption{ A sketch of the recurrent ResNet network model used for evaluation.
  The network consists of 24 residual blocks, RRA blocks added after blocks 3,6,9,12,15 and 24, and skip connections after the first 4 RA blocks.
  The internal dimension of all convolutional layers is 80, except for the kernel predictive layers, which are $5\times5\times5=125$ large.}
  \Description{RA-ResNet diagram.}
  \label{fig:resnet}
\end{figure*}

\section{Related Work}
\textbf{Kernel-predictive neural networks}
Our work is closely related to \cite{Bako2017,Vogels2018,Zhang2021,Zhang2023}, since we use kernel-predictive denoising networks, i.e., the output of the network are per-pixel 2D (or 3D) kernels.
Instead of predicting the colors the denoised image directly, the kernels are applied (via dot product) on the input image to denoise it.
The main difference between our method and these related works is in the neural network components and loss for temporal denoising, but the remaining techniques are compatible with our method.
Mildenhall et al.~\shortcite{Mildenhall2018} employ a kernel predictive network for denoising burst photographs (instead of rendered images).
Their loss generalization is similar to ours in that it introduces loss terms for the individual images in the burst.
We instead pair images before and after the central frame to combat occlusion and disocclusion, and do not downscale temporal terms while training.
Instead of filtering the image after it is rendered, sample-based methods for error reduction such as \cite{Gharbi2019} correct pixel samples before aggregating them into a framebuffer.
We scope this paper on using temporal information for denoising, and leave out discussion on techniques that run as the individual frames are rendered like \cite{Cho2021,Yu2021}.

\textbf{Recurrent neural networks for denoising}
Introducing recurrent connections in denoising neural networks has been done, for example by Chaitanya et al.~\shortcite{Chaitanya2017} and Hasselgren et al.~\shortcite{Hasselgren2020}.
The purpose of these connections is the same as in our work: to force the network to use temporal information when denoising.
Our approach differs in that it is bidirectional, i.e., we take advantage of succeeding as well as previous frames.
The remaining differences with these works are due to the target applications.
We trade off network complexity to achieve higher image quality.

\textbf{Non-neural denoising}
Denoising techniques which do not rely on neural networks can also be used in production.
Methods like Non-Local Means Filtering \cite{Buades2005} or Guided Filtering \cite{He2013}, use model-based approaches to derive per-pixel weights used for spatial denoising.
Other similar techniques were extended to include temporal information \cite{Dabov2007} in addition to the larger body of work for film restoration which can be repurposed for Monte Carlo denoising \cite{Kokaram2014}.
We use such a denoiser to generate pseudo-reference images, but leave further discussion on non-neural denoising outside the scope of this paper.
Finally, we recommend the survey on adaptive sampling and denoising by Zwicker et al.~\shortcite{Zwicker2015} for the readers looking for an overview of the field.

\section{Robust Average Blocks}
\begin{figure}[ht]
  \centering
  \includegraphics[width=\linewidth]{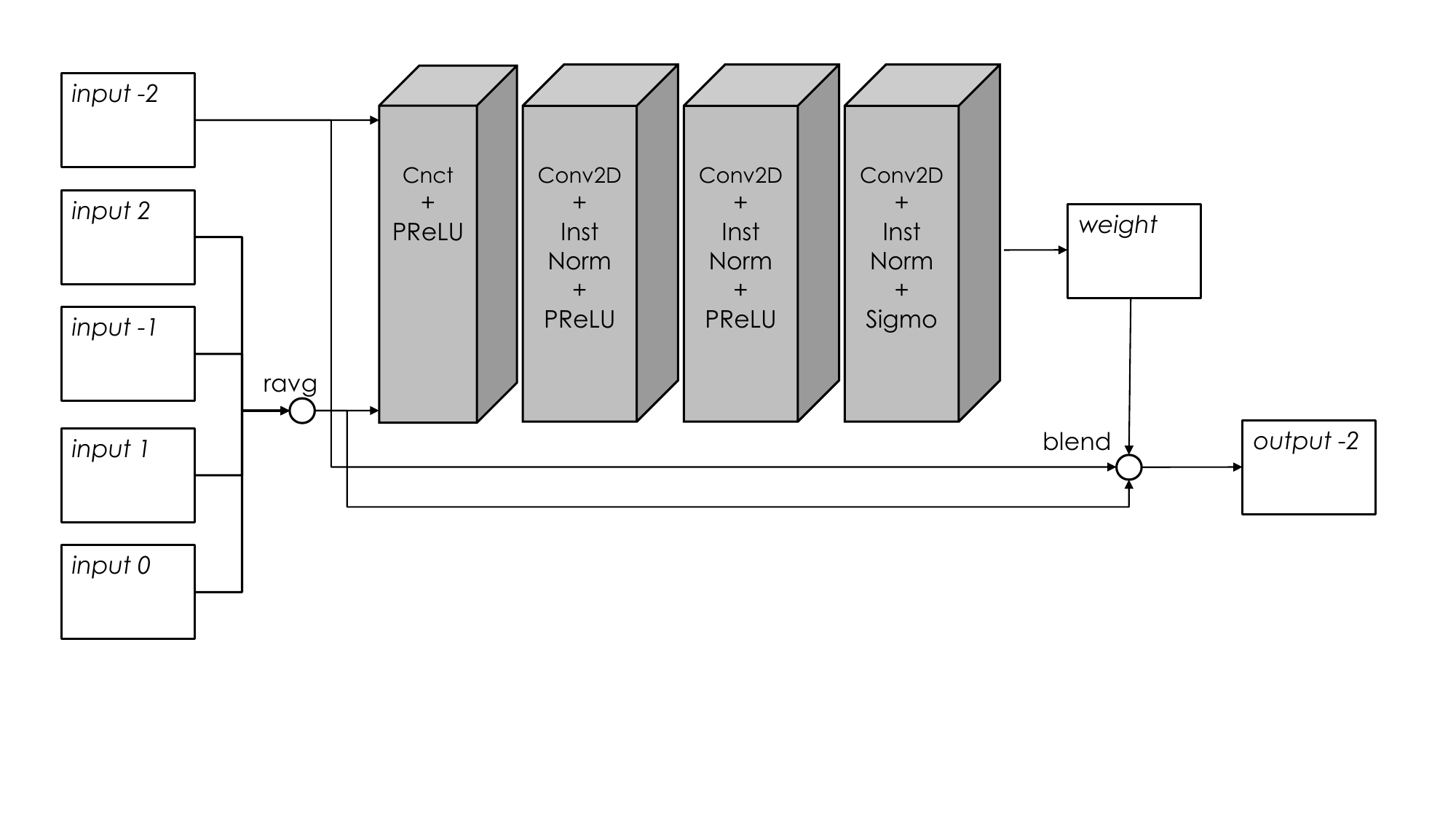}
  \caption{Robust Average block for a sequence of length 5.
  We exclude the first frame, average the remaining frames, and interpolate between the average and the excluded frame.
  This is repeated (recurrently) until each frame of the sequence has been interpolated with the robust average of the other frames.
  % In this example, we start from the outermost frames and proceed inwards ending with the central frame.
  }
  \Description{Network layer diagram for RA blocks.}
  \label{fig:rra}
\end{figure}
In the following, we introduce the Robust Average (RA) blocks.
We use them to force the network to denoise in time via latent space interpolation.
Intuitively, averaging the internal state across several noisy estimates of the same image will result in similar network output and reduce temporal noise.
However, we did not obtain satisfactory results using a straight forward implementation that just averages network state at various depths.
The networks we tried were prone to creating artifacts (e.g. light leaks).
Our understanding is that averaging in latent space reduced the ability of the network to remove outliers caused by noise or warping errors.

We address this problem in two ways.
First, we let the network independently predict interpolation weights and use them to blend each frame of the sequence with an average of the remaining frames.
This enables outlier removal and preserves the ability of the network to represent an identity transformation.
The latter is important for networks like ResNet \cite{He2016}.
% In the case of augmenting a spatial network, reverting to identity inside the RA blocks makes the resulting network spatial instead of temporal. 
Second, given enough frames, we use an averaging operation that is robust to outliers: given a set of values $S \subset \mathbb{R}$ with $|S| \geq 3$, we call \emph{robust average} computing the mean after discarding the minimum and maximum:
\begin{equation}
  ravg(S) = \frac{1}{|S| - 2}\left(\sum_{S}{x} - \max(S) - \min(S)\right)
\end{equation}
% This operation is useful for outlier removal when combining random samples.
% In the following, we adopt its working principle for temporal denoising.

% We convert spatial convolutional neural networks, e.g. ResNet \cite{He2016}, into temporal ones by adding Robust Average (RA) blocks (Figure \ref{fig:resnet}) at various depths.
% These blocks perform interpolation (and outlier removal) in the time dimension and operate on the network's internal state.

% The Robust Average block shown in Figure \ref{fig:rra} works in the following way.
The Robust Average block shown in Figure \ref{fig:rra} operates on a fixed length window of frames centered at the current frame.
First, we exclude one frame from the sequence, and compute robust average of the remaining frames.
Then a set of convolutional layers is used to estimate per-pixel weights, which are then used to blend the excluded frame with the average.
% The result is then moved to the end of the sequence as if the sequence starts from the following frame for the next iteration.
This is repeated once for each frame in the sequence.
Note that the operation of the RA block is recurrent since each successive step uses the result of the previous step.

The main advantage of the RA blocks compared to averaging, is that the balance between using temporal and spatial information for each pixel is learned instead of fixed.
The network can reduce or increase the amount of temporal interpolation per pixel at different depths.
In addition to noise reduction, this helps with correcting errors due to inaccurate warping.

% The recurrent robust average blocks perform interpolation in latent space and force the network to denoise in time as well as in space.
% Network parameters in the remaining, spatial, components are reused across the sequence of frames.
% Importantly, the amount of influence of each frame is learned, because the network predicts the interpolation weights (Figure \ref{fig:rra}).
% Note however that the model can revert to a static one by training to not interpolate across frames in the sequence, i.e. identity remains in the set of functions the network can represent.

% The parameters of all layers in the RA block are shared across the entire sequence of frames.
% Instead of relying entirely on the network to find corresponding pixels in the input frames, we pre-warp the input to convert the images, as close as possible, to estimates of the same frame.
We use motion vectors generated during rendering to warp frames on each side of the central frame.
This includes auxiliary output buffers such as the albedo and normals.
We also compute confidence in the warp from the motion vectors the warped albedo and normal buffers.
In order to avoid temporal artifacts without decreasing weights along temporal connections, we mix in pixel values from the central frame inversely proportional to the confidence score for each side frame.
This helps prevent the network from degenerating to a spatial only model during training.

The number of frames used as input is fixed because the blocks are bidirectional.
However, we are able to extend the temporal range by running multiple denoising passes.
This is possible for networks with high detail preservation and low bias such as sharpening.

As the blocks can be added at any depth, they allow filtering in time in the very first layers.
This is in contrast with approaches that denoise in space first, and only then introduce temporal connections.
In the evaluation, we show evidence that our method helps retain small scale image features that are blurred by a network with a deep spatial component followed by a temporal part.
Furthermore, adding RA blocks at multiple depths reduces the size and expressive power of the spatial-only components of the network.
This favors denoising in the time dimension, and hinders the model from degenerating into a spatial-only during training.

\section{Thresholded Kernel Prediction}

A common choice for activation for the output layer for kernel-predictive networks is the softmax function:
\begin{equation}
  \sigma(w_i) = \frac{e^{w_{i}}}{\sum_{j} e^{w_{j}}}
\end{equation}
This has the advantage that the weights $w_{i}$ are normalized to sum up to 1, however the predicted weights are strictly positive.
With non-zero kernel weights, every pixel in the kernel support contributes to the final estimate.
We found that this can lead to artifacts when high intensity pixels are present in the kernel support.
The high intensity pixels can be fireflies, lightsources, or specular reflections of lightsources, and they amplify inaccuracies in the predicted weights.

For intuition, consider a value $y \in \mathbb{R}$ that is reconstructed from a set of estimates $x_i$, using kernel weights $w_i$.
If there are weights $w_j = 0$ participating in the reconstruction, it holds that $y$ can be reconstructed without additional error even if $x_j$ becomes an outlier, e.g., a firefly.
Note that in the case of Monte Carlo denoising, the objective is to reconstruct the correct pixel values from random estimates.
It is therefore likely that the estimates for the same pixel regions sometimes contain fireflies, i.e., pixels with comparatively larger error.

We enable zero kernel weights by using use thresholded ReLUs followed by normalization in the kernel outputting layer of our networks.
More specifically, our activations are:
\begin{equation}
  \eta_t(w_i) = \frac{\max(0, w_i - t)}{\sum_{j=1}^K \max(0, w_j - t)}
\end{equation}
with $t < \frac{1}{K}$ to prevent zeroing out normalized kernels with equal weights throughout the domain.
If not mentioned otherwise, we report results using $t = \frac{1}{2K}$.
If all weights are below $t$, we replace the resulting kernel with identity.
Note that thresholds can be used with any activation, including softmax, at the expense of an additional normalization if it is desirable for the kernel weights to sum up to 1.

\begin{figure}[ht]
  \centering
  \def\svgwidth{\linewidth}
  
  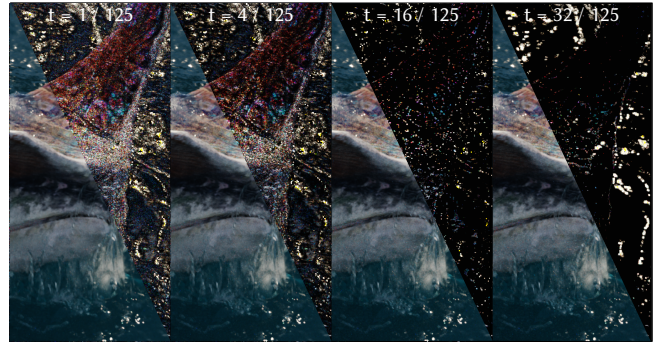
  \caption{ Increasing the value of the kernel threshold $t$ adjusts the influence of the denoiser on the image.
  The figure shows denoised images and scaled difference to the noisy input image. 
  \textcopyright 20th Century Studios / Walt Disney Studios Motion Pictures}
  \Description{Specular denoising.}
  \label{fig:threshold}
\end{figure}

Increasing the threshold can be used to correct loss of detail in denoised images at the expense of reintroducing noise from the input (see Figure \ref{fig:threshold}).
This accommodates fine-tuning the model after training based on artist feedback, if done once before the trained model is deployed.
Alternatively, the threshold value can be user controlled and adjusted per image.
Finally, note that in our case increasing the kernel threshold usually results in less temporal utilization.
This is because for each weight for the central frame, we have multiple weights for remaining frames.

\section{Spatial-to-Temporal Loss Conversion}
\label{sec:loss}

Denoising via convolutional neural networks is commonly formulated as follows.
Let $X := \{x^{1}, \dots, x^{N}\}$ and $Y:=\{y^{1}, \dots, y^{N}\}$ be two corresponding sets of noisy and reference images, and $l(\cdot,\cdot) : \mathbb{R}^n \times \mathbb{R}^n \rightarrow \mathbb{R}$ a per-pixel image loss (e.g. mean absolute error).
Let $f_\theta(\cdot) : \mathbb{R}^n \rightarrow \mathbb{R}^n$, be a convolutional neural network parameterized by $\theta$ that denoises one image at a time.
The optimal set of parameters for the network is determined via supervised learning:
\begin{equation}
  \argmin_{\theta}\frac{1}{N}\sum^N_{i=1} l\displaystyle\left(f_\theta(x^i), y^i\displaystyle\right)
\end{equation}
In the following, we set $N = 1$ to simplify notation.

We propose to generalize the procedure from a spatial into a temporal one by asking the neural network to predict the reference image from any subset of the input image sequence.
Let $X := \{x_{-k}, \dots, x_{k}\}$ be a set of consecutive noisy frames and $y_0$ the reference for the center frame.
Assume that for each $i \in [-k,k]$, there is a motion compensating transformation (image warp) that aligns features from $x_i$ to $x_0$, and let $X':=\{x'_{-k}, \dots x'_k\}$ be the set of transformed images.
Consider a family of neural networks $G_\theta :=\{g^i_\theta(\cdot)\}$ such that for each $\hat{X} \subset X'$, there is a $g^i_\theta(\hat{X}) : \cdot \rightarrow \mathbb{R}^n$.
The minimization problem then becomes:
\begin{equation}
  \argmin_{\theta}\sum_{\hat{X} \in \mathcal{P}(X')} l\displaystyle\left(g^i_\theta(\hat{X}), y_0\displaystyle\right)
\end{equation}
where $\mathcal{P}(X')$ is the power set of $X'$.
Given $g_\theta(\cdot) : \mathbb{R}^{n(2k+1)} \rightarrow \mathbb{R}^n$, a spatio-temporal kernel-predictive neural network for $X'$, a set $G_\theta$ is easy to construct by zeroing out predicted kernel weights for the excluded frames and re-normalizing the remaining weights.
In other words, we assume that the neural networks predict an image from a set of estimates instead of a single one.
We treat noise, warping error, and other biases as deviations from the target image and rely on the network to correct for them.

\begin{figure}[ht]
  \centering
  \includegraphics[width=\linewidth]{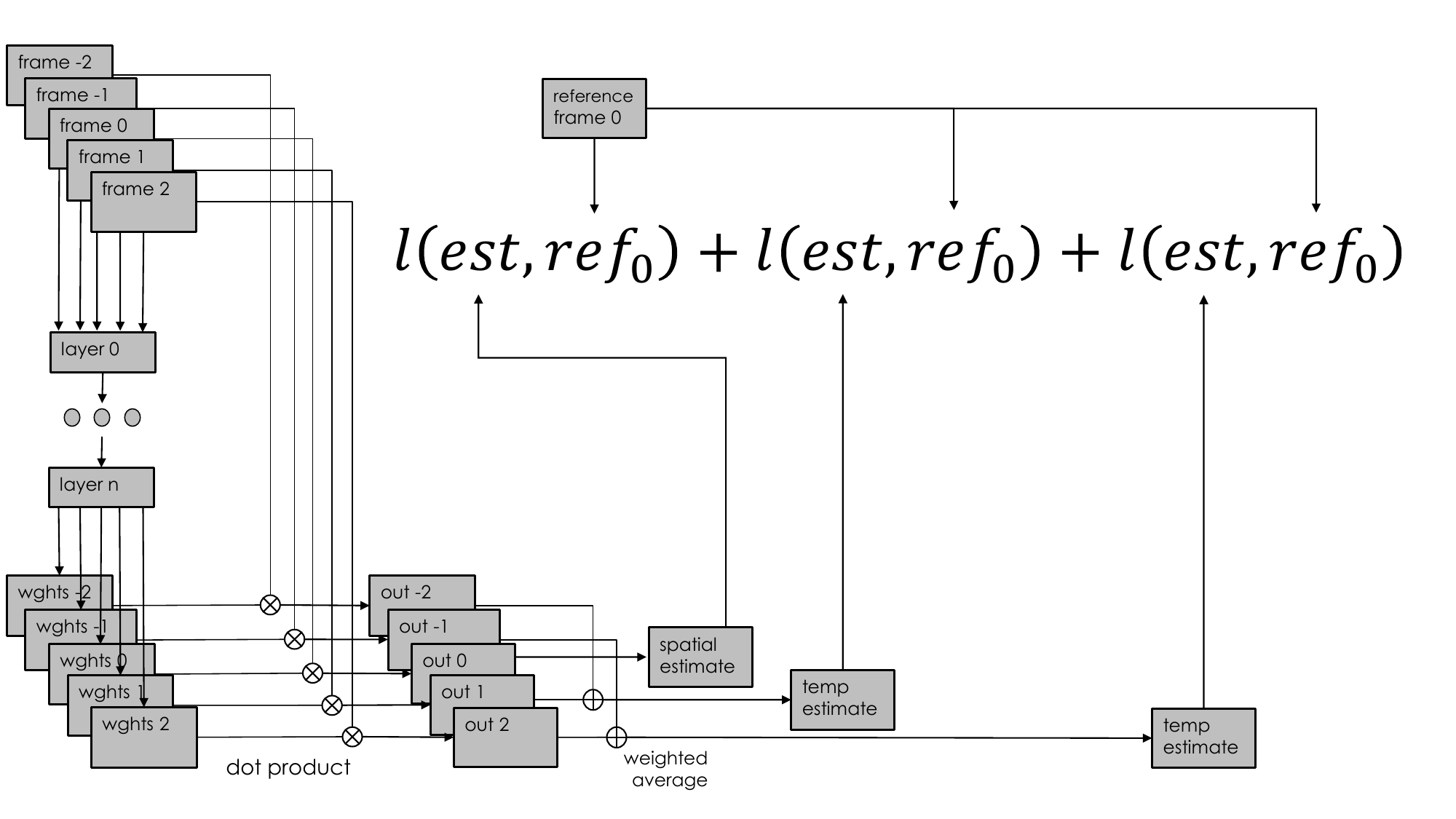}
  \caption{
  Spatial to temporal loss conversion for a kernel-predictive neural network.
  Here, we add two temporal loss terms that force the network to predict the center frame from pairs of one previous and one subsequent frame.
  We use the same reference frame $ref_0$ in each term.
  }
  \Description{Conversion of spatial losses to temporal ones.}
  \label{fig:loss}
\end{figure}

In practice, we do not construct a network and loss term for each subset of the input sequence to reduce memory and computational complexity.
Instead, as shown in Figure \ref{fig:loss}, we train a network with temporal window of 5 with loss terms for:
\begin{itemize}
  \item the center frame
  \item a weighted average of the first and fourth frame
  \item a weighted average of the second and fifth frame
\end{itemize}
we compare against the center reference frame in each case.
We select pairs of frames before and after the center frame, to improve the chance of retaining image features that are occluded after or revealed during the central frame.
We only use one term containing the center input frame in order to promote the use of temporal information.

Note that our approach works with a spatial loss function instead of depending on measured deviations in time.
Because of that, we are able to train without temporal reference images, while still penalizing error using the center frame.
As a result, the network learns to reduce error coming from the off-center frames that is caused by either noise, motion, or warping.
Temporal artifacts can be further penalized by adding a loss term that compares the combined result to the reference frame.
Note that Vogels et al.~\shortcite{Vogels2018} train the temporal component of their denoiser using just the combined loss term.
In our case: end to end training of a spatio-temporal network, this steered the network to use less temporal information.
Therefore, we only used a downweighted global loss term to post-train models if they introduce temporal artifacts.

Our conversion works by adding spatial loss terms for output images generated via temporal averaging, independent of the choice of spatial loss $l(\cdot, \cdot)$.
We tested this by training with various losses including mean absolute error ($L_1$), symmetric mean absolute percentage error \cite{Vogels2018}, and the perceptual loss in \cite{Johnson2016,Gatys2016}.
We did not observe substantial differences in the learning behavior in the different cases.
However, networks trained with a perceptual loss had better detail preservation compared to the remaining losses.
We therefore train using

\begin{equation}
  l(x, y) := \displaystyle\left(L_{VGG_{5,4}}(x, y) + L_{VGG_{3,4}}(x, y)\displaystyle\right)L_{SMAPE}(x,y)
\end{equation}
where $L_{VGG_{a,b}}(\cdot, \cdot)$ is a $L_1$ distance in the latent space of block $a$, layer $b$ of VGG19 \cite{VGG19} trained on ImageNet \cite{imagenet}.
The choice of addition and multiplication for the separate losses is to take into account the implicit scaling of the terms.
We sum terms with similar magnitude and multiply otherwise.
We opted for SMAPE instead of $L_{VGG_{2,2}}(\cdot, \cdot)$ as a "feature" loss, because the latter led to aliasing artifacts.
A simpler $L_1$ loss can be used instead of SMAPE depending on color space and pixel value range and distribution.

\section{Training Data}

Creating large bodies of training data for machine learning methods can be challenging in a VFX environment because of the effort required to produce imagery.
It is often the case that generating a fully converged reference images or image sequences for denoising is prohibitively expensive.
Based on the results demonstrated by Lehtinen et al.~\shortcite{Lehtinen2018}, we do not train using ground truth reference images.
Instead, we render images at different noise levels and use them to construct input and target pairs for training.
In addition to the noisy references we also use pseudo-references generated by an existing denoiser.
The latter can be a previous version of the denoising network, or in our case, a separate method based on using a modified TV-L1 \cite{Wedel2009} optical flow method initialization from the renderer-provided motion vectors, confidence based robust averaging, Non-Local Means Filtering \cite{Buades2005}, and Guided Filtering \cite{He2013}.

The training data we use for evaluation has been generated by instrumenting a production path tracer to render images at 3 noise levels:
\begin{itemize}
  \item \emph{noisy} RGB image using $n$ samples per pixel
  \item \emph{half quality} RGB image using $\left\lfloor\frac{n}{e}\right\rfloor$ samples per pixel\footnote[1]{We use an irrational fraction to ensure that subsequences of quasi-random samples remain low discrepancy.}
  \item \emph{low quality} RGB image using 4 samples per pixel
\end{itemize}
The number of samples taken varies per image and per pixel.
The noise level of the images is typically low enough to deliver final quality after denoising and compositing.
During training, we pick the pairs of noise levels randomly (including pseudo-references) and denoise from more noisy to less noisy.

We train on data from approximately 11000 rendered sequences of length of up to several hundred frames.
From these, we randomly sample 10 disjoint sets of 16000 to 20000 tiles of dimensions $128\times128\times11$.
%The split into sets is to reduce time for each epoch, which allows more frequent learning rate adjustments and progress monitoring.

\begin{figure}[ht]
  \centering
  \def\svgwidth{\linewidth}
  
  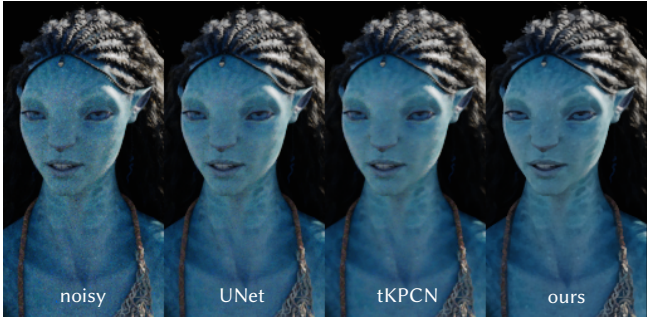
  \caption{
  Our denoiser delivers similar image quality with slightly better image details compared to a baseline (tKPCN) network without RA blocks.
  A spatial UNet keeps detail by preserving noise, and sacrificing temporal stability.
  \textcopyright 20th Century Studios / Walt Disney Studios Motion Pictures}
  \Description{Temporal breakdown.}
  \label{fig:dn_compare}
\end{figure}

\section{Evaluation}
\begin{figure}[ht]
  \centering
  \def\svgwidth{\linewidth}
  
  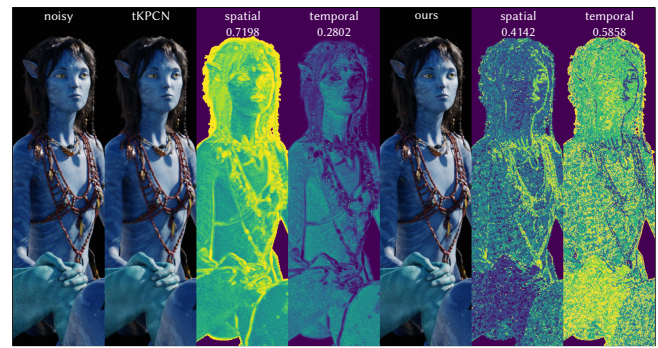
  \caption{ Spatial vs temporal contributions for our denoiser compared to a baseline (tKPCN) using a temporal loss and network like \cite{Vogels2018} instead of our loss and RA blocks.
  \textcopyright 20th Century Studios / Walt Disney Studios Motion Pictures}
  \Description{Temporal breakdown.}
  \label{fig:tmp_spt}
\end{figure}

In this section, we compare our network extension via RA blocks to an alternative spatio-temporal architecture used in \cite{Vogels2018} which is, to our knowledge the most similar alternative in terms of target application and network structure.
We show that it is possible to obtain at least similar results for a single image (Figure \ref{fig:dn_compare}, Table \ref{tab:psnr}), while using larger temporal weights (Figure \ref{fig:tmp_spt}).
Note that our networks do not contain temporal connections other than the ones in the RA blocks (even in the final kernel-predictive layers).
The network being able to balance weights between the central and non-central frames is thanks to these components.

\begin{figure}[th]
  \centering
  \def\svgwidth{\linewidth}
  
  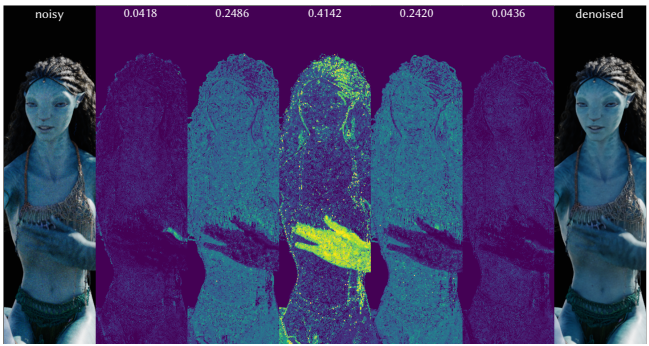
  \caption{ Temporal breakdown of denoising weights. Noisy, denoised image, and contributions for the input sequence of 5 frames.
  The outermost frames have significantly smaller total contribution, however the minimum and maximum weights for each frame are 0 and 1.
  \textcopyright 20th Century Studios / Walt Disney Studios Motion Pictures}
  \Description{Temporal breakdown.}
  \label{fig:tmp_weights_2118}
\end{figure}

The Robust Average blocks can be used together with a number of convolutional neural network types.
We tested this on ResNet and UNet architectures, as well as the multiscale ResNet variant in \cite{Vogels2018}.
The trained models performed sufficiently well to be used for noise reduction in a commercial rendering pipeline.
The key contributing factors for this were the absence of introduced temporal flickering and the detail/edge preservation (see Figure \ref{fig:dn_compare}).
Compared to noise reduction via sampling, the denoised static images compared favorably to frames with $\frac{1}{4}$ estimated pixel error.
For final quality renders, we were not able to achieve lower temporal noise than a denoised sequence of frames by increasing pixel samples.

From the topologies we tested, the best performing network is a ResNet (Figure \ref{fig:resnet}) without upscaling and downscaling components.
The network denoises from a 5 frame temporal window using $5 \times 5 \times 5$ kernels (height, width, time), and has 24 million trainable parameters.
The input for the network is pre-warped using motion vectors and consists of color, albedo, normals, and confidence.
The latter is a single channel image representing confidence in the warp derived from the motion vectors. 

Figure \ref{fig:tmp_weights_2118} is an example from a scene with relatively slow camera or character movement.
The average pixel contribution from the central frame is smaller than the combined weight of the remaining frames, meaning that the denoiser uses more temporal than spatial information.
The outermost frames have significantly smaller average contribution.
This is expected, given that in a large random training set the temporal coherency and image warp quality decreases fast with the distance to the central frame.
On the other hand, the minimum and maximum weights for each frame are 0 and 1.
In other words, the network can completely replace pixel values from the central frame from the surrounding frames.

When compared against a baseline, the RA blocks and temporal loss improve temporal utilization.
We remove all RA blocks from the network, and instead use the temporal module and loss proposed in \cite{Vogels2018}.
As with their work, the spatial parameters are still shared across the sequence of input frames.
This network, which we call tKPCN, can train to denoise in time, however strongly favors the center frame if trained with a standard loss.
Figure \ref{fig:tmp_spt} shows a crop from the same scene as Figure \ref{fig:tmp_weights_2118}, which exhibits good temporal coherency.
The baseline method uses smaller weights for pixels from the non-center frames both on average: (0.0063, 0.1158, 0.1503, 0.007) compared to (0.0418, 0.2486, 0.2420, 0.0436);
and in terms of maximum weights: (0.1163, 0.9636, 0.8337, 0.1122) compared to 1.

\begin{table}[t]
	\begin{footnotesize}
	\begin{center}
	\resizebox{\columnwidth}{!}{
	\begin{tabular}{rrrrrrr}
	%\hline
	              &   	&   		  &   		  &         &   	    & frame id\\
	              &   	& -2		  & -1		  &   0     &  1	    & 2		    \\
	\cmidrule(lr){3-7}
	%\hline
  RA +         & avg & 0.0197  & 0.1671  & 0.6409  & 0.1514  & 0.0209  \\
  spatial loss  & max & 0.2013  & 0.3223  & 0.9966  & 0.3199  & 0.1950  \\
                &     &         &         &         &         &         \\
	RA +         & avg & 0.0580  & 0.1402  & 0.5694  & 0.1531  & 0.0792  \\
  our loss      & max & 0.9959  & 0.9982  & 1.0000	& 0.9925  & 0.9582  \\
                &     &         &         &         &         &         \\
	tKPCN +       & avg & 0.0021  & 0.1387  & 0.7200  & 0.1379  & 0.0013  \\
  spatial loss  & max & 0.1505  & 0.9956  & 1.0000  & 0.9975  & 0.3985  \\
                &     &         &         &         &         &         \\
  tKPCN +       & avg & 0.0387  & 0.2994  & 0.4076  & 0.2123  & 0.0421  \\
  our loss      & max & 0.9831  & 0.9999  & 1.0000  & 0.9987  & 0.9964  \\
                &     &         &         &         &         &         \\

  %\hline
	\end{tabular}
	}%resize box
	\end{center}
	\end{footnotesize}
\caption{\label{tab:rra_ablation} Ablation tests on a frame from the Payakan scene (see Figure \ref{fig:nn_cmp}).
RA blocks provide better temporal utilization with a spatial-only loss compared to the non-RA network trained for 11 epochs (35k training samples each).
Both networks converge to using predominantly spatial information with further training with the spatial loss.
Our loss reformulation does improve temporal utilization for the tKPCN network.}
\end{table}

\begin{figure}[ht]
  \centering
  \def\svgwidth{\linewidth}
  
  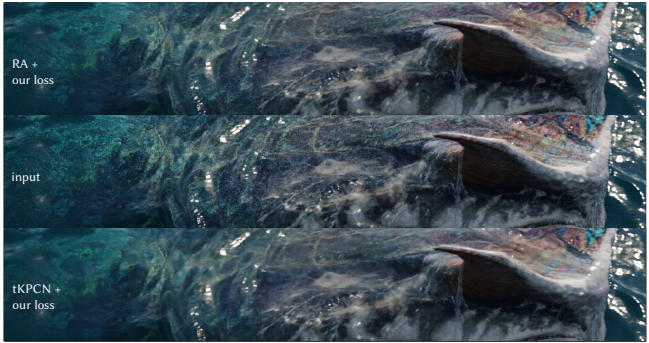
  \caption{ Image comparison using our network and a tKPCN model trained with our temporal loss reformulation (from Table \ref{tab:rra_ablation}).
  The tKPCN model removes more noise, while our network preserves more detail, especially on parts of the image where the creature is submerged underwater.
  \textcopyright 20th Century Studios / Walt Disney Studios Motion Pictures}
  \Description{Model comparison.}
  \label{fig:nn_cmp}
\end{figure}

We break down the influence of the network structure and the loss function in Table \ref{tab:rra_ablation}.
Both our network and the tKPCN baseline can be trained to favor the central frame when using a "spatial" loss term measuring the difference to the reference image.
This is not surprising, since there is no incentive for the network to use the pixel values from the surrounding frames.
The RA blocks slow down the rate at which the network converges to a mostly spatial network.
On the other hand, the temporal loss reformulation can steer both our network and the baseline to use larger temporal weights, however we were not able to train a tKPCN network with a good temporal utilization, no artifacts, and better detail preservation compared to our RA models (see Figure \ref{fig:nn_cmp}).
In our experience, it was easier to train a tKPCN model to remove more noise, while RA networks are better at edge preservation.
For this comparison, we trained the both networks with identical losses and procedure: an initial pass with the temporal loss followed by pass with added downscaled spatial loss term.
We report results using the checkpoints with the best validation loss for both networks.

The test in Figure \ref{fig:tmp_weights_2118} shows the network predicting predominantly spatial weights for the hand of the character, which moves fast, and predominantly temporal weights on the character face and body which remain mostly static.
Reverting to spatial denoising in the absence of temporal coherency reduces or avoids image artifacts such as ghosting and light leakage which occur when interpolating between non-corresponding pixel values in time.

\begin{figure}[ht]
  \centering
  \def\svgwidth{\linewidth}
  
  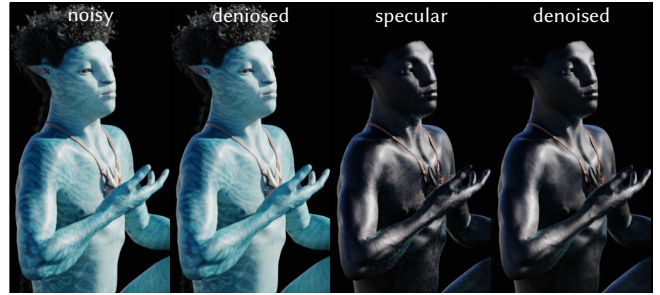
  \caption{ Denoising the specular component of a render with the kernel weights from the RGB output can work well in practice.
  Some small details in the specular component of the image are still recognizable after denoising.
  \textcopyright 20th Century Studios / Walt Disney Studios Motion Pictures}
  \Description{Specular denoising.}
  \label{fig:spec}
\end{figure}

Figure \ref{fig:spec} shows the level of detail we are able to obtain when predicting weights from an image and applying them only to the specular component of the image.
While not mathematically correct, this works well enough in practice and recuces computational effort when denoising renders with a lot of auxiliary buffers.

In Table \ref{tab:psnr} we compare peak signal-to-noise ratio (PSNR) and structural similarity index measure (SSIM) on our datasets.
We aggregate statistics from a set of test images that have not been used for training or validation.
In the test we denoise from the half quality RGB image to a pseudo-reference denoised with a production denoiser.
In VFX, the quality of the denoised images is judged by expert users, which makes the assessment perceptual rather than statistical.
The test results here do match our assessment of the relative image quality: both models deliver similar results, and our network is consistently better at keeping image detail.

\begin{table}[ht]
	\begin{footnotesize}
	\begin{center}
	\resizebox{\columnwidth}{!}{
	\begin{tabular}{rrrrrrrr}
	%\hline
            &    	    &   		  & PSNR	  &     &   	    &   		  & SSIM		\\
			      & min	    & avg		  & max		  &     & min	    & avg		  & max			\\
	\cmidrule(lr){2-4} \cmidrule(lr){6-8}
	%\hline
	input	    & 42.4898 & 44.6739 & 46.8580 & 	  & 0.9880  & 0.9900  & 0.9920  \\
	tKPCN     & 42.0127 & 44.9809 & 47.9492 &     & 0.9877  & 0.9910  & 0.9944  \\
  ours      & 45.3110 & 48.0516 & 50.7922 &     & 0.9946  & 0.9957  & 0.9969  \\
            &         &         &         &     &         &         &         \\
  %\hline
	\end{tabular}
	}%resize box
	\end{center}
	\end{footnotesize}
\caption{\label{tab:psnr} Aggregate SSIM and PSNR comparison between our model and the temporal baseline network.
The statistics are from 768 test samples from scenes that did not participate in training.}
\end{table}

\section{Limitations}

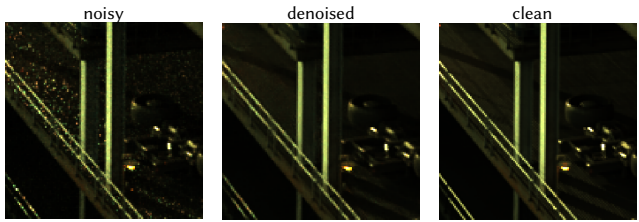
\begin{figure}[ht]
  \centering
  \def\svgwidth{\linewidth}
  
  %% Creator: Inkscape 1.3 (0e150ed6c4, 2023-07-21), www.inkscape.org
%% Accompanies image file 'test_beauty_low.pdf' (pdf, eps, ps)
\begingroup%
  \makeatletter%
  \providecommand\color[2][]{%
    \errmessage{(Inkscape) Color is used for the text in Inkscape, but the package 'color.sty' is not loaded}%
    \renewcommand\color[2][]{}%
  }%
  \providecommand\transparent[1]{%
    \errmessage{(Inkscape) Transparency is used (non-zero) for the text in Inkscape, but the package 'transparent.sty' is not loaded}%
    \renewcommand\transparent[1]{}%
  }%
  \providecommand\rotatebox[2]{#2}%
  \newcommand*\fsize{\dimexpr\f@size pt\relax}%
  \newcommand*\lineheight[1]{\fontsize{\fsize}{#1\fsize}\selectfont}%
  \ifx\svgwidth\undefined%
    \setlength{\unitlength}{826.63998413bp}%
    \ifx\svgscale\undefined%
      \relax%
    \else%
      \setlength{\unitlength}{\unitlength * \real{\svgscale}}%
    \fi%
  \else%
    \setlength{\unitlength}{\svgwidth}%
  \fi%
  \global\let\svgwidth\undefined%
  \global\let\svgscale\undefined%
  \makeatother%
  \begin{picture}(1,0.3422687)%
    \lineheight{1}%
    \setlength\tabcolsep{0pt}%
    \put(0,0){\includegraphics[width=\unitlength,page=1]{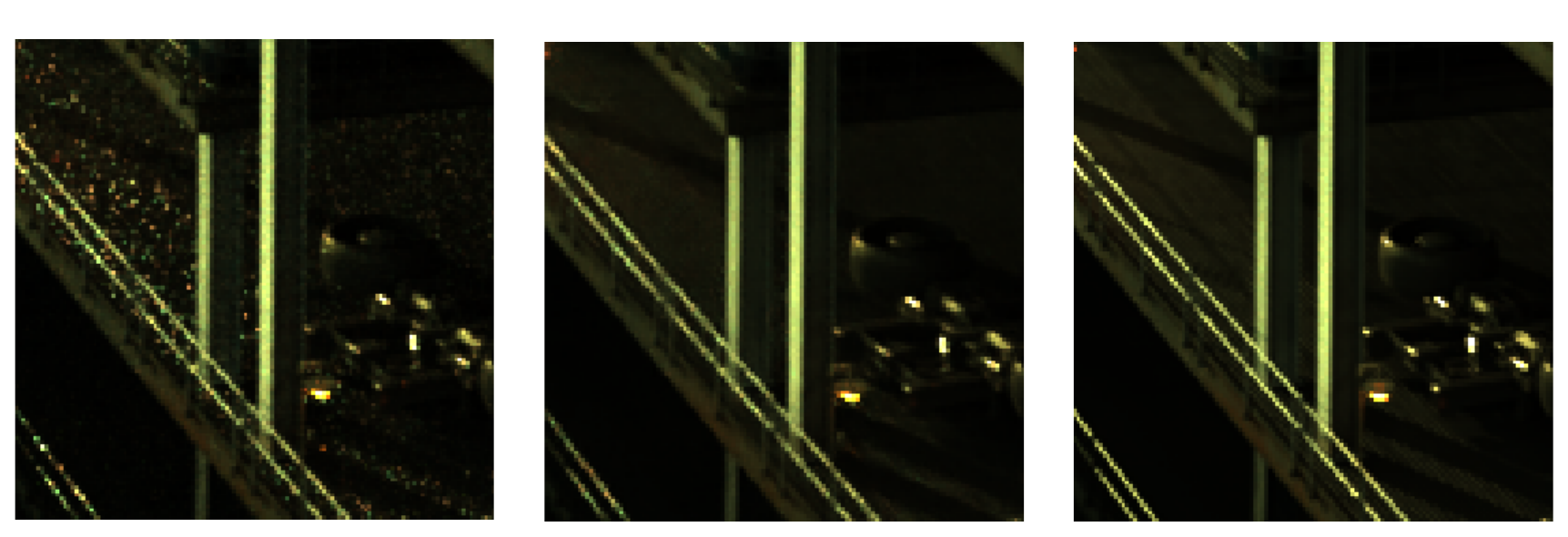}}%
    \put(0.16222386,0.32216661){\color[rgb]{0,0,0}\makebox(0,0)[t]{\lineheight{1.25}\smash{\begin{tabular}[t]{c}\footnotesize{\textsf{noisy}}\end{tabular}}}}%
    \put(0.44740686,0.32216661){\color[rgb]{0,0,0}\makebox(0,0)[lt]{\lineheight{1.25}\smash{\begin{tabular}[t]{l}\footnotesize{\textsf{denoised}}\end{tabular}}}}%
    \put(0.79809806,0.32174536){\color[rgb]{0,0,0}\makebox(0,0)[lt]{\lineheight{1.25}\smash{\begin{tabular}[t]{l}\footnotesize{\textsf{clean}}\end{tabular}}}}%
    \put(-0.20582605,0.5018136){\color[rgb]{0,0,0}\makebox(0,0)[lt]{\begin{minipage}{1.49018167\unitlength}\centering \end{minipage}}}%
  \end{picture}%
\endgroup%

  \caption{ Validation image with low sample count.
  Our network fails to reconstruct small scale shadows present in the pseudo-reference image.
  }
  \Description{High noise example.}
\label{fig:beauty_low}
\end{figure}

We show a reconstruction from a low sample count input in Figure \ref{fig:beauty_low}.
While the reconstruction is plausible, details like the shadow of the railing are missing.
In this case, the network is not able to predict image detail revealed by further sampling.
With our data and training, we favor preserving detail and sacrifice the ability to reconstruct from very low initial sample counts.
Note that the reconstruction quality from lower sample counts can be accetable for certain use cases, e.g., previews.

The main disadvantage of extending networks via Robust Average blocks is the increased network graph complexity.
Mixing in temporal components at different network depths makes it hard to train spatial and temporal parts of the network separately.
In our case, we had to compensate for this by training with batch size of 1 or 2 on a GPU with 24GB of memory.
Memory restrictions can also hinder increasing the temporal window of the neural network.
Our results suggest that there is not much potential in expanding beyond 5 frames without improving the warping quality.
We instead increase the temporal window by running multiple denoising passes if necessary.

The large network graph also impacts inference time negatively.
Our implementation in Python using Tensorflow 2.5 \cite{tensorflow2015} needs approximately 2 min to denoise a 2k stereo image on a high-end GPU (Nvidia RTX A5000).
This includes loading the network from a file and compiling the network graph, warping the input frames and auxiliary buffers, kernel prediction on tiles of size $728 \times 548$, kernel thresholding, warping output buffers, and kernel application.
In practice, we amortize the network initialization by denoising in batches of frames, and optimize for efficiency by running on CPUs with low thread counts, which results in longer inference times.

Finally, denoising in time assumes temporal coherency in the input sequence.
If coherency in time is not present, we rely on the network to revert to using spatial information.
Our networks can, for example, preserve shadows moving over static surfaces, however we currently use a spatial denoiser for renders of fire and explosions.

\section{Conclusion}
In this paper, we introduce a method for designing and training temporal denoising neural networks.
We introduced a new recurrent block pattern, an improved training technique, and an alternate method for using temporal information in image sequences.
We believe that our work helps make a small step towards higher image fidelity and rendering efficiency in VFX,
and hope that this paper will help and inspire further research in the field.
\begin{acks}
The authors thank Mihail Moskov for consulting on deep learning and image processing.
We thank Joe Letteri, Sam Cole, and Francois Sugny for guidance on imaging and compositing, and Youngbin Park for pipeline integration. 
This work has received a Technical Achievement Award from the Academy of Motion Picture Arts and Sciences.
\end{acks}

\bibliographystyle{ACM-Reference-Format}
\bibliography{bibliography}

\appendix

%\section{Appendix}


\end{document}